\documentclass[conference]{IEEEtran}
\IEEEoverridecommandlockouts

\usepackage{cite}
\usepackage{amsmath,amssymb,amsfonts}
\usepackage{algorithmic}
\usepackage{graphicx}
\usepackage{textcomp}
\usepackage{xcolor}
\usepackage{graphicx}
\usepackage{cite}
\usepackage{amsmath,amssymb,amsfonts}
\usepackage{algorithmic}
\usepackage{graphicx}
\usepackage{textcomp}
\usepackage{xcolor}
\usepackage{comment}
\usepackage{hyphenat}
\usepackage{multirow}
\usepackage{tabu}
\usepackage{caption}
\usepackage{subcaption}
\usepackage{amsthm}
\theoremstyle{definition}
\newtheorem{definition}{Definition}
\usepackage{multicol, blindtext}
\usepackage{breqn}
\usepackage{
tikz,
relsize,
amsmath,
booktabs,
tikz
}
\usetikzlibrary{arrows,calc,positioning,shadows,shapes}
\usepackage{pgfplots}
\usepackage{pgfplotstable}
\pgfplotsset{compat=1.6}

\usepackage[T1]{fontenc}
\usepackage[utf8]{inputenc}
\usepackage{lmodern}

\usepackage[
labelfont=sf,
hypcap=false,
format=hang,
width=0.8\columnwidth
]{caption}

\usetikzlibrary{
calc,trees,shadows,positioning,arrows,chains,
decorations.pathreplacing,
decorations.pathmorphing,
decorations.shapes,
decorations.text,
shapes,
shapes.geometric,
shapes.symbols,
matrix,
patterns,
intersections,
fit}

\pgfdeclarelayer{background layer}
\pgfdeclarelayer{foreground layer}
\pgfsetlayers{background layer,main,foreground layer}

\tikzset{
>=latex
}

\usepackage[utf8]{inputenc}
\usepackage{amsmath}
\usepackage{amssymb}
\usepackage{arydshln}
\usepackage{mathrsfs}
\usepackage[export]{adjustbox}
\usepackage{float}
\usepackage{times}
\usepackage{fancyhdr,graphicx,amsmath,amssymb}
\usepackage[ruled,vlined]{algorithm2e}
\newtheorem{theorem}{Theorem}
\include{pythonlisting}

\def\BibTeX{{\rm B\kern-.05em{\sc i\kern-.025em b}\kern-.08em
    T\kern-.1667em\lower.7ex\hbox{E}\kern-.125emX}}

\makeatletter
\newcommand*\titleheader[1]{\gdef\@titleheader{#1}}
\AtBeginDocument{%
  \let\st@red@title\@title
  \def\@title{%
    \bgroup\normalfont\large\centering\@titleheader\par\egroup
    \vskip1.5em\st@red@title}
}

\def\ps@IEEEtitlepagestyle{%
  \def\@oddfoot{\mycopyrightnotice}%
  \def\@evenfoot{}%
}
\def\mycopyrightnotice{%
  {\footnotesize  978-1-6654-6837-4/22/\$31.00 \copyright 2022 IEEE \hfill}
  \gdef\mycopyrightnotice{}
}

\makeatother
\usepackage{lipsum}
\usepackage{xcolor}
\usepackage{soul}
\sethlcolor{black}
\makeatletter
\newif\if@blind
\@blindtrue 
\if@blind \sethlcolor{black}\else
   
\fi

\begin{document}





\title{Towards Direct Comparison of Community Structures in Social Networks}

 \author{\IEEEauthorblockN{Soumita Das\textsuperscript{1}, Anupam Biswas\textsuperscript{2}}\\
 \IEEEauthorblockA{\textit{Department of Computer Science and Engineering} \\
 \textit{National Institute of Technology Silchar}\\
 Silchar, Assam, India \\
 Email: \textsuperscript{1}wingsoffire72@gmail.com, \textsuperscript{2}anupam@cse.nits.ac.in}
 }


\maketitle


\begin{abstract}

Community detection algorithms are in general evaluated by comparing evaluation metric values for the communities obtained with different algorithms. The evaluation metrics that are used for measuring quality of the communities incorporate the topological information of entities like connectivity of the nodes within or outside the communities. However, while comparing the metric values it loses direct involvement of topological information of the communities in the comparison process. In this paper, a direct comparison approach is proposed where topological information of the communities obtained with two algorithms are compared directly. A quality measure namely \emph{Topological Variance (TV)} is designed based on direct comparison of topological information of the communities. Considering the newly designed quality measure, two ranking schemes are developed. The efficacy of proposed quality metric as well as the ranking scheme is studied with eight widely used real-world datasets and six community detection algorithms.

\end{abstract}

\begin{IEEEkeywords}
Community detection Algorithms, Node, Social Network, Community Evaluation, Quality, Performance
\end{IEEEkeywords}

\section{Introduction}
The exponential growth of Social Networks (SN) has given a thrust to the research community to explore and analyze the properties of these networks. Social entities such as users tend to get associated with other social entities having similar interests~\cite{bedi2016community,fottovat2022community}. These behavior of the social entities give rise to interesting connectivity patterns in social networks termed as communities. The formation of communities are driven by the interaction and closeness of entities and hence, the relationship between entities play a significant role in the quality of communities~\cite{abiswas2015}.  Detection of such community structures helps in understanding the relationships between the entities in SN. The primary objective of community detection is to extract a group of densely connected entities present in a network where entities within the group are strongly connected and entities across groups are weakly connected~\cite{girvan2002community}. 

Detection of communities has been in the forefront of social network analysis in recent years. Numerous community detection algorithms have been developed, which require proper evaluation in order to determine their performance. Community detection algorithms in general are evaluated from the perspective of accuracy and quality. Since real-world networks mostly do not have the ground truth, evaluation process has to rely solely on quality aspect where primarily topological information of communities are considered. Though, measuring quality does not require ground truth, due to lack of standard definition of a community poses another challenge in the evaluation process as different quality measures are being designed to apprehend different aspects of communities. Most importantly, the comparison of metric values is also not reliable since communities obtained with two different algorithms may have completely different topology and still they can have same metric value. Moreover, comparison of metric value does not exhibit direct comparison of ground realities of the entities of network that belong to different communities. Therefore, it is quite logical to compare topological information of entities like connectivity of the nodes within or outside the communities directly instead of comparing metric values for the communities obtained with different algorithms.

In this paper, a direct comparison approach is proposed where topological information of entities are considered for comparing communities directly. In general, entities in social network are referred to both nodes and edges. We have considered nodes and its surrounding topology to analyze it's impact on the quality of communities obtained with two different algorithms. Communities listed in two community sets are paired if they share maximum number of common nodes. Nodes in each community pair are examined for their surrounding topological differences and thereby accounting all such difference a quality measure is designed. The main contributions of this work are listed as follows:

\begin{itemize}


  \item A node level quality measure namely, Topological Variance ($TV$) have been introduced, which accounts the topological difference of nodes' surrounding with respect to community pairs from the two community sets obtained with two different algorithms.

   
   \item Two ranking schemes are proposed considering the newly designed quality measure to generate Dataset wise Topological Variance ($DTV$) and Overall Topological Variance ($OTV$) for an algorithm in comparison to others.
   

   
  
  \item An extensive empirical analysis is performed on eight real-world networks having varied statistics and six popular community detection algorithms.
  
  \item Performed comparative analysis of community detection algorithms in terms of  using $TV$ score, $DTV$ and $OTV$ ranking.
   
\end{itemize}

The rest of the paper is organized as follows: Section~\ref{three} discusses about the proposed direct comparison approach, Section~\ref{four} demonstrates about our experimental setup, followed by Section~\ref{five} which discusses the results and finally the conclusion in Section~\ref{six}.


\section{Proposed Approach}
\label{three}

This section details about the proposed quality evaluation measure for community evaluation and ranking schemes. The traditional method of evaluating communities is by using evaluation metrics which is an indirect way of evaluating communities. This is indirect in the sense that although the structure of communities are considered by the evaluation metrics but structure of the communities are not evaluated with respect to each other. Rather the scores obtained from the evaluation metrics are used for evaluation purpose. Here, we have proposed an evaluation measure to evaluate community structures directly with respect to each other. Here, direct evaluation is carried by examining and comparing two community structures with respect to each other. Let us now try to understand our proposed approach by initializing all the prerequisites and discussing some of the preliminary concepts followed by our proposed measure and ranking schemes.


We have considered an undirected, unweighted graph $G(V,E)$ where $V$ indicates set of nodes and $E$ indicates set of edges in the graph $G$.  In this paper, we have referred the algorithms that are used for performance comparison as primary algorithm and alternative algorithm. Primary algorithm $A^{p}$ is the algorithm whose performance is to be evaluated. Also, the algorithm with respect to which primary algorithm is compared is called alternative algorithm $A^{a}$. Here, the performance of these algorithms are evaluated considering the communities given by primary algorithm and alternative algorithm referred as primary communities and alternative communities respectively. The number of communities given by these algorithms may be equal in number or may not be equal in number. Therefore, we shall use the symbols $k$ and $l$ to indicate the number of communities given by primary algorithm and alternative algorithm represented by $C^{p}=\{C_{1}^{p},C_{2}^{p},C_{3}^{p},...,C_{k}^{p}\}$ and $C^{a}=\{C_{1}^{a},C_{2}^{a},C_{3}^{a},...,C_{l}^{a}\}$ respectively, where, $k$ is not necessarily equal to $l$. Now, pairing the communities given by  $C^{p}$ and $C^{a}$ is necessary for comparative analysis, but such pairing cannot be done randomly. It is because the comparison of two community list ($C^{p},C^{a}$) requires pairing of similar communities. These type of pairing that are based on similarity of the communities are defined as comparable communities which are detailed below. 


\begin{definition}(Comparable communities). Two communities $C_{i}^{p}$ and $C_{j}^{a}$ from primary and alternative communities respectively are comparable communities when $\underset{j}{\operatorname*{arg\,max}} ~~C_i^p  \cap C_j^a $.
\end{definition}

 
 \begin{definition}(Analytical nodes). An analytical node is a node that is used for finding the topological difference of comparable community pair $(C_{i}^{p},C_{j}^{a})$ . All such nodes that are present in $C_{i}^{p}$ only but not in $C_{j}^{a}$ are analytical nodes when $C_{i}^{p} - C_{j}^{a}$.
 \end{definition}
 
It is important to find how the comparable community pair $(C_{i}^{p},C_{j}^{a})$ vary in terms of topology given by Topological Variance for comparable community pair. Then, the summation of  Topological Variance for all comparable community pairs gives the Topological Variance for primary community and alternative community under consideration. The Topological Variance is detailed below.

\subsection{Topological Variance} 
In this section, we shall discuss about a novel measure called Topological Variance which gives the topological difference of primary communities and alternative communities. Here, topological difference is used to determine the performance of primary communities with respect to alternative communities. Basically, the computation of of Topological Variance comprises of two steps. Firstly, Topological Variance of each of the comparable community pair is computed. Secondly, combination of Topological Variance of all comparable community pairs gives the Topological Variance of primary communities and alternative communities. These two aspects are detailed below.




\begin{definition}{(Topological Variance of comparable community pair).}  Topological Variance of comparable community pair ($C_{i}^{p},C_{j}^{a}$) is used to find the topological difference of $C_{i}^{p}$ and $C_{j}^{a}$ by utilizing the topological information of analytical nodes associated with $(C_{i}^{p},C_{j}^{a})$ pair. Topological Variance of comparable community pair ($C_{i}^{p},C_{j}^{a}$) is defined as,


\begin{align}
 \label{eone}
 \footnotesize
  TV(C_{i}^{p},C_{j}^{a})=&\frac{1}{\mid C_{i}^{p}-C_{j}^{a} \mid} \times \footnotesize \sum \left(\frac{\Gamma(v_{i}) \cap C_{i}^{p}}{\mid \Gamma(v_{i})\mid} ~~~~~~~~\right.\nonumber\\
  &\left. - \frac{\Gamma(v_{i}) \cap C_{j}^{a}}{\mid \Gamma(v_{i})\mid}\right), \forall v_{i} \in C_{i}^{p}, v_{i} \notin C_{j}^{a}, 
\end{align}
\end{definition}

where, node $v_{i}$ is an analytical node used for analyzing topological variance of $C_{i}^{p}$ and $C_{j}^{a}$.  The topological variance is divided by the total number of analytical nodes to normalize the measure. The term $\Gamma(v_{i})$ refers to the set of neighbors of node $v_{i}$ and $\mid \Gamma(v_{i})\mid$ indicates total number of neighbors of  $v_{i}$. Next, topological variance for two community lists $C^{p}$ and $C^{a}$ belonging to primary communities and alternative algorithms is obtained by using Topological Variance for comparable community pairs.


\begin{definition}{(Topological Variance).}
The Topological Variance of primary communities and alternative communities is used to find the topological difference of these two communities. Here, the topological difference is obtained by the summation of Topological Variance of all comparable community pairs. Topological Variance for primary communities $C^{p}$ and alternative communities $C^{a}$ is defined by,


\begin{align}
   \label{etwo}
   \footnotesize
TV(C^{p},C^{a})=&\frac{1}{\mid \Delta(C^{p},C^{a}) \mid} \times \nonumber \\ \footnotesize ~~~~~~~~ &\sum_{\substack{\forall{(C_{i}^{p},C_{j}^{a})  \in  \Delta(C^{p},C^{a})}}}{TV(C_{i}^{p},C_{j}^{a})}
\end{align}
\end{definition}

where, the normalized Topological Variance value is obtained by dividing the summation term by total number of analytical nodes indicated by $\mid \Delta(C^{p},C^{a}) \mid$. When the value of Topological Variance is greater than zero, it indicates that primary communities shares greater number of connections than the number of connections in alternative communities and hence, the quality of primary communities is better than alternative communities. Whereas, when the value is less than zero, it indicates that alternative communities possess greater number of connections as compared to primary communities  and therefore, the quality of  alternative communities is better than primary communities and value equal to zero indicates that both primary and alternative communities share almost same number of connections and hence, they possess almost similar quality communities.

Node $v_{i}$ in equation~\ref{eone} is a analytical node because it is present in $C_{i}^{p}$ but not in $C_{j}^{a}$ where $(C_{i}^{p},C_{j}^{a})$ is a comparable community pair. The summation of the difference of the number of connections that all such $v_{i}$ nodes shares with $C_{i}^{p}$ and $C_{j}^{a}$ respectively is indicated by the second term. This term is divided by the total  number of nodes that are present in $C_{i}^{p}$ only, indicated by $\mid C_{i}^{p}-C_{j}^{a} \mid$. The term $\Gamma(v_{i})$ refers to the set of neighbors of node $v_{i}$ and $\mid \Gamma(v_{i})\mid$ indicates total number of neighbors of  $v_{i}$. Next,  summation of the $TV$ score of all the comparable community pairs is divided by the total number of comparable community pairs indicated by $\mid \Delta(C^{p},C^{a}) \mid$ to obtain the $TV$ score which is defined by,

\begin{align}
   \label{etwo}
   \footnotesize
TV(C^{p},C^{a})=\frac{1}{\mid \Delta(C^{p},C^{a}) \mid} \times \nonumber \\ \footnotesize ~~~~~~~~ \sum_{\substack{\forall{(C_{i}^{p},C_{j}^{a}) \in \Delta(C^{p},C^{a})}}}{TV(C_{i}^{p},C_{j}^{a})}
\end{align}

\begin{theorem}
Range of $TV$ is [-1,1].
\end{theorem}

\begin{proof}

The definition of Topological Variance for comparable community pair is defined in Eq.~\ref{eone}. In this equation, the summation term indicates the number of connections that an analytical node share with $C_{i}^{p}$ and $C_{j}^{a}$. Here, an analytical node may share all it's neighbors with $C_{i}^{p}$, then the first term of summation term becomes 1 and when some of the neighbors of an analytical node do not belong to  $C_{i}^{p}$, then in the worst case scenario, minimum value of this term is greater than 0 but less than 1. Whereas, in most of the cases, an analytical node do not share all it's neighbors with $C_{j}^{a}$ and therefore, maximum value of the second term associated with summation term is less than 1 but greater than 0 and the minimum value is 0. Considering these maximum and minimum terms and normalizing by total number of analytical nodes gives Topological Variance for comparable community pair to be 1, less than 1 or greater than -1. Next, summation of the Topological Variance of all comparable community pairs and dividing by total number of such pairs gives maximum Topogical Variance to be less than 1 and minimum value as greater than -1. 
\end{proof}


The performance of the primary algorithms with respect to the alternative algorithms needs to be evaluated. Two ranking schemes are developed based on Topological Variance and are discussed in the next section. 

\begin{table}[htbp] 
\caption {Dataset Statistics.}
\label{dat}
\centering
 \begin{tabular} {| m{6em}|m{6em}|m{6em}|m{6em}|}
 \hline
 \centering{Dataset} & \centering{\# ~Nodes} & \centering{ \# ~Edges} & \centering{Average degree}
 \tabularnewline [4pt]
\hline
  \centering{Riskmap~\cite{riskmap}} & \centering{42} &  \centering{83}  &  \centering{3.95} 
  \tabularnewline[4pt]
  \centering{Football~\cite{football}} & \centering{115} &  \centering{613}&  \centering{10.66}
  \tabularnewline [4pt]
  \centering{Polblogs~\cite{polblogs}} & \centering{1222} &  \centering{16718}  &  \centering{27.31} 
  \tabularnewline[4pt]
  \centering{Jazz~\cite{jazz}} & \centering{198} &  \centering{2742}  &  \centering{27.70} 
 \tabularnewline[4pt]
  \centering{Dolphin~\cite{dolphin}} & \centering{62} &  \centering{159}  &  \centering{5.12} 
  \tabularnewline[4pt]
 \centering{Polbooks~\cite{polbooks}} & \centering{105} &  \centering{441}  &  \centering{8.4}
  \tabularnewline[4pt]
  \centering{Strike~\cite{strike}} & \centering{24} &  \centering{34}  &  \centering{3.16} 
  \tabularnewline [4pt]
 \centering{Sawmill~\cite{sawmill}} & \centering{36} &  \centering{37}  &  \centering{3.44} 
  \tabularnewline [4pt]
   \hline
   \end{tabular}
\end{table}

\subsection{Performance Comparison}
It is important to compare the performance of community detection algorithms to understand which algorithm gives the best quality community. There are three ways of carrying comparative performance such as one-to-one comparison, one-to-many comparison and many-to-many comparison. In this paper, we have proposed a measure called Topological Variance based on one-to-one comparison. Using this measure, we have proposed two types of ranking schemes i.e.  Dataset wise Topological Variance ($DTV$) and Overall Topological Variance $(OTV)$ which are based on a many-to-many comparison. $DTV$ ranking is based on specific dataset, $OTV$ ranking gives an overall idea about the performance of primary algorithm on all the datasets. To discuss these ranking schemes, firstly, we have introduced Average Topogical Variance  $(ATV)$ which gives the average performance of a primary algorithm with respect to alternative algorithms considering Topological Variance. Suppose, we have $m$ number of algorithms such as $A_{h}=A_{1},A_{2},..,A_{m}$. At a time, one of these algorithms will be considered as primary algorithm. Average Topogical Variance for $i^{th}$ algorithm $A_{i}$ acting as primary algorithm  is defined as,

\begin{align}
    \label{avgssr}
    ATV(A_{i}^{p})=\frac{1}{|m|}\sum_{h=1}^{ m}TV(C^{p},C^{a})
\end{align}

 where, primary communities $C^{p}$ is given by $A_{i}$ and alternative communities $C^{a}$ is given by all the $m$ algorithms. The performance of $C^{p}$ is based on the value of Average Topological Variance. High Average Topological Variance indicates good performance of primary algorithm. This strategy is later used for ranking based on specific datasets and ranking based on overall performance on all datasets.
 

\textbf{Dataset wise Topological Variance ($DTV$).} The performance of primary algorithms on specific datasets is evaluated incorporating Dataset wise Topological Variance ranking scheme. Algorithm with maximum  Average Topological Variance is ranked as 1 (highest rank).  \\

\textbf{Overall Topological Variance ($OTV$).}
The performance of primary algorithms with respect to all the alternative algorithms on all datasets is determined by Overall Topological Variance ranking scheme. The Overall Topological Variance ranking for a primary algorithm $A_{i}^{p}$ is defined as,\\


\begin{align}
    \label{ors}
    OTV(A_{i}^{p})= desc(\frac{1}{\mid D \mid}(\sum_{d=1}^{ D }ATV(A_{h}^{p}))), \nonumber \\ \forall A_{h},~~ where, h=1,..,m.
\end{align}

where, $desc$ indicates descending order of arrangement, such that maximum $ATV$ value on all the datasets gives the Overall Topological Variance rank as 1.

\begin{table*}[ht!]
\captionsetup{width=2\columnwidth}
\caption { $TV$ values related to a primary algorithm with respect to alternative algorithms on all datasets.}
\label{tvv}
\centering
  \begin{tabular} {| m{12em}| m{6.5em} | m{6.5em} | m{5.5em}|m{6.5em}|m{6.5em}|m{6.5em}|}
   \hline
   \multirow{3}{*}{\centering{Primary Algorithm}}&  \multirow{3}{*}{Dataset}&
   \multicolumn{5}{c|}{Alternative algorithms}\\
   \cline{3-7}
   &  & \multirow{2}{*}{SCAN}  &  \multirow{2}{*}{LPA}  & \multirow{2}{*}{GM}  & \multirow{2}{*}{Gdmp2} & \multirow{2}{*}{Kcut}  \\
    &&&&&&\\
   \hline
   \centering{}  &\centering{Riskmap} & \centering{0.948} & \centering{0.956} &\centering{0.956} & \centering{0.968} & \centering{0.641} 
   \tabularnewline
  \cline{2-7}
   \centering{}  &\centering{Football} & \centering{0.626} & \centering{0.517} &\centering{0.579} & \centering{0.629} & \centering{-0.270} 
   \tabularnewline
  \cline{2-7}
   \centering{AGDL}  &\centering{Polblogs} & \centering{0.499
} & \centering{0.303} &\centering{0.464} & \centering{0.996} & \centering{0} 
   \tabularnewline
  \cline{2-7}
   \centering{}  &\centering{Jazz} & \centering{0.875} & \centering{0.864} &\centering{0.890} & \centering{0.979} & \centering{0.132} 
   \tabularnewline
  \cline{2-7}
   \centering{}  &\centering{Dolphin} & \centering{0.649} & \centering{0.363} &\centering{0.437} & \centering{0.672} & \centering{-0.135} 
   \tabularnewline
  \cline{2-7}
   \centering{}  &\centering{Polbooks} & \centering{0.599} & \centering{0.581} &\centering{0.434} & \centering{0.602} & \centering{0.086} 
   \tabularnewline
  \cline{2-7}
   \centering{}  &\centering{Strike} & \centering{0.992} & \centering{0.887} &\centering{0.958} & \centering{0.926} & \centering{0.633} 
   \tabularnewline
  \cline{2-7}
   \centering{}  &\centering{Sawmill} & \centering{0.990} & \centering{0.819} &\centering{0.974} & \centering{0.968} & \centering{0.196} 
   \tabularnewline
  \hline
 \end{tabular}
\end{table*}

\section{Experimental Setup}
\label{four}

Experiments are conducted considering three aspects such as, community detection algorithms, datasets and evaluation metrics. Specifically, we have selected algorithms that considers the neighborhood information of vertices, network structure or modularity optimization. For e.g. Structural Clustering Algorithm for Networks $(SCAN)$, $Gdmp2$, $AGDL$ are based on neighborhood information of vertices. Whereas,  Label Propagation Algorithm $(LPA)$ considers network structure, Greedy-modularity $(GM)$ and $Kcut$ are based on modularity maximization~\cite{clauset2004finding,xu2007scan, raghavan2007near, blondel2008fast,cem,chen2010dense,zhang2012graph,ruan2007efficient}. Moreover, several real-world small and medium scale datasets have been selected having varied average degree to check the performance of these community detection algorithms. We have chosen real-world datasets having ground truths only inorder to verify the communities given by the baseline algorithms with the respective ground truths and then evaluate the quality. The following datasets have been used for our experimental purpose:\\

\textbf{Datasets}: Riskmap, Football, Polblogs, Jazz, Dolphin, Polbooks, Strike and Sawmill obtained from publicly available online repositories such as, SNAP~\cite{snapnets} and network repositories~\cite{networkrepsitories}. The details of each of these datasets are summarized in Table~\ref{dat}.\\

\textbf{Evaluation metrics:} Several evaluation metrics  based on internal and external connections have been selected  to perform a comparative analysis of our proposed measure,  Topological Variance. The following evaluation metrics have been used for our experiments: Modularity based on internal connections, Conductance based on external connections, Isolability based on combination of internal and external connections~\cite{mfca,biswas2017defining}.

\begin{figure*}[htbp]
\centering
\centering {\includegraphics[width =18.2cm]{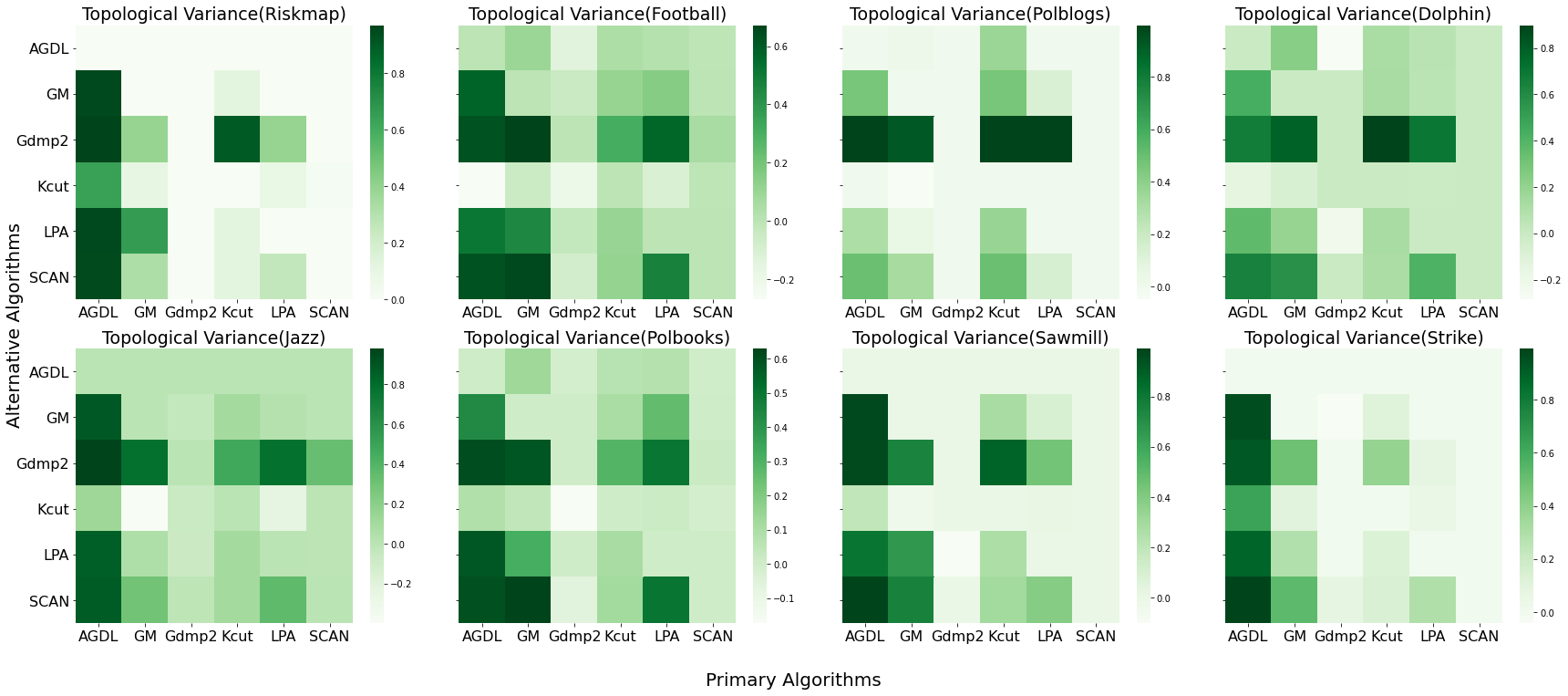}}
\captionsetup{width=1.9\columnwidth}
\caption{\footnotesize Eight different heat maps showing comparative performance of primary algorithm with respect to alternative algorithm on eight datasets respectively. X-axis represents primary algorithms, Y-axis represents alternative algorithms, each of the cells in a heatmap represents the $TV$ score of primary algorithm with respect to alternative algorithm.}

\label{ELA}
\end{figure*}


\pgfplotstableread[row sep=\\,col sep=&]{
$DTV$ & SCAN & LPA & GM & Gdmp2 & AGDL & Kcut\\
Riskmap & 5 &  4  & 3 & 6 & 1 & 2 \\
Football & 5& 3  & 2 & 6 & 1 & 4\\
Polblogs  &	5 & 4 &	3& 5 & 2 & 1\\
Jazz & 5& 2 & 4 & 6 & 1 & 3 \\
Dolphin& 5	& 4  &	2 & 6 & 1& 3\\
Polbooks & 5 & 3 & 2 & 6 & 1 & 4\\
Strike & 6 & 4 & 2 & 5 & 1 & 3 \\
Sawmill  & 5& 4  &	3 & 6  & 1  & 2\\
}\lookm
\begin{figure*}[h!]
\centering
\begin{tikzpicture}
    \begin{axis}[
            ybar,
            bar width=.206cm,
            width=1.0001\textwidth,
            height=.3\textwidth,
            enlarge x limits=0.10,
            legend style={at={(0.5, .975)},
                anchor=north,legend columns=6,legend cell align=left},
            symbolic x coords={Riskmap,Football,Polblogs,Jazz,Dolphin,Polbooks,Strike,Sawmill },
            xtick=data,
             x tick label style={rotate=00,anchor=north},
            nodes near coords align={vertical},
            ymin=0,ymax=8.2,
            ylabel={$DTV$},
            nodes near coords,
            every node near coord/.append style={rotate=90, anchor=west}
        ]
        \addplot table[x=$DTV$,y=SCAN]{\lookm};
        \addplot table[x=$DTV$,y=LPA]{\lookm};
        \addplot table[x=$DTV$,y=GM]{\lookm};
        \addplot table[x=$DTV$,y=Gdmp2]{\lookm};
        \addplot table[x=$DTV$,y=AGDL]{\lookm};
        \addplot table[x=$DTV$,y=Kcut]{\lookm};
        \legend{SCAN,LPA,GM,Gdmp2,AGDL,Kcut}
    \end{axis}
\end{tikzpicture}
\captionsetup{margin=2cm}
\caption{ Comparative analysis of community detection algorithms based on $DTV$ ranking.} 
\label{quah}
\end{figure*}
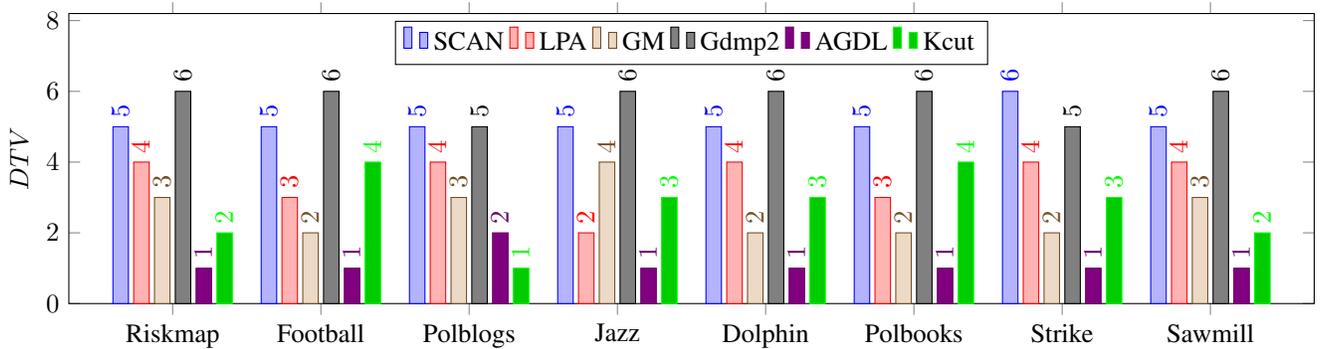

\section{Result Analysis}
\label{five}

\begin{table*}[htbp] 
\captionsetup{width=2\columnwidth}
\caption{ Overall comparative ranking of primary algorithm with all the alternative algorithms on all the datasets indicated by $OTV$.}


\label{elr}
\centering
 \begin{tabular} {| m{8.0em} |c|c|c|c|c|c|c|c|c|c|}
 \hline
 \centering{ } & &\centering{Riskmap} & \centering{Football} & \centering{Polblogs} & \centering{Jazz} & \centering{Dolphin}&\centering{Polbooks} & \centering{Strike}& \centering{Sawmill}&\centering{\small $OTV(A_{i}^{p})$}
 \tabularnewline
 \cline{2-11}
  \centering{$ATV(A_{i}^{p})$} & \centering{SCAN} &  \centering{0.003}  &  \centering{0.009} &   \centering{0} &   \centering{0.054} &   \centering{0} &   \centering{0.001} &   \centering{0} &   \centering{0} &  \centering{5} 
  \tabularnewline [1pt]
 & LPA &  0.123  &  0.186 &   0.211 &   0.157 &   0.204 &   0.225 &   0.071 &   0.159 &   \centering{4}
  \tabularnewline [1pt]
  \centering{} & \centering{GM} &  \centering{0.148}  &  \centering{0.304} &   \centering{0.212} &   \centering{0.123} &   \centering{0.288} &   \centering{0.285} &   \centering{0.236} &   \centering{0.356} &   \centering{2}
  \tabularnewline  [1pt]
  \centering{} & \centering{Gdmp2} &  \centering{0}  &  \centering{-0.079} &   \centering{0} &   \centering{-0.031} &   \centering{-0.084} &   \centering{-0.04} &   \centering{0.006} &   \centering{-0.017} &   \centering{6}
 \tabularnewline  [1pt]
  \centering{} & \centering{AGDL} &  \centering{0.745}  &  \centering{0.347} &   \centering{0.377} &   \centering{0.623} &   \centering{0.331} &   \centering{0.384} &   \centering{0.733} &   \centering{0.658} &   \centering{1} 
  \tabularnewline  [1pt]
 \centering{} & \centering{Kcut} &  \centering{0.213}  &  \centering{0.114} &   \centering{0.448} &   \centering{0.131} &   \centering{0.232} &   \centering{0.116} &   \centering{0.126} &   \centering{0.289} &   \centering{3}
  \tabularnewline  [1pt]
  \hline
   \end{tabular}
\end{table*}

\begin{table*}[ht!]
\captionsetup{width=2\columnwidth}
\caption {Results of each of community detection algorithms with several evaluation metrics.}
\label{rem}
\centering
  \begin{tabular} {| m{4em}| m{4em}| m{5em}| m{5em}|m{5em}|m{4em}|m{5em}|m{5em}|m{5em}|}
   \hline
   \multirow{3}{*}{\centering{Algorithm}}&  \multirow{3}{*}{Dataset}&
   \multicolumn{3}{c|}{Evaluation metrics} & \multirow{3}{*}{Dataset}&
   \multicolumn{3}{c|}{Evaluation metrics}\\
   \cline{3-5}
   \cline{7-9}
   &  & \multirow{2}{*}{Isolability}  &  \multirow{2}{*}{Modularity}  & \multirow{2}{*}{Conductance} & &  \multirow{2}{*}{Isolability}  &  \multirow{2}{*}{Modularity}  & \multirow{2}{*}{Conductance} \\
    &&&&&&&&\\
   \hline
   \centering{SCAN}  &\centering{} & \centering{0.050} & \centering{0.029} &\centering{0.6} & \centering{} & \centering{0.086}& \centering{0.067} & \centering{0.575} 
   \tabularnewline
  \cline{1-1}
  \cline{3-5}
  \cline{7-9}
  \centering{LPA}  &\centering{} & \centering{0.045} & \centering{0.498
} &\centering{0.352
} & \centering{} & \centering{0.094}& \centering{0.481} & \centering{0.229} 
   \tabularnewline
 \cline{1-1}
  \cline{3-5}
  \cline{7-9}
  \centering{GM}  &\centering{Dolphin} & \centering{0.065} & \centering{0.495} &\centering{0.316} & \centering{Polbooks} & \centering{0.108 }& \centering{0.481
} & \centering{0.229
} 
   \tabularnewline
 \cline{1-1}
  \cline{3-5}
  \cline{7-9}
  \centering{Gdmp2}  &\centering{} & \centering{0.0} & \centering{0.004} &\centering{0.875} & \centering{} & \centering{0.02}& \centering{0.040} & \centering{0.786} 
   \tabularnewline
 \cline{1-1}
  \cline{3-5}
  \cline{7-9}
  \centering{AGDL}  &\centering{} & \centering{0.5374} & \centering{0.408} &\centering{0.260} & \centering{} & \centering{0.622}& \centering{0.044} & \centering{0.275} 
   \tabularnewline
  \cline{1-1}
  \cline{3-5}
  \cline{7-9}
  \centering{Kcut}  &\centering{} & \centering{0.049} & \centering{0.021} &\centering{0.890} & \centering{} & \centering{0.062}& \centering{0.010} & \centering{0.887} 
   \tabularnewline
  \hline
 \end{tabular}
\end{table*}
The communities obtained with various community detection algorithms on selected datasets $D$ have been examined and compared with our proposed measure namely, \textsl{Topological Variance}. The comparative results have been discussed below.

\subsection{Result analysis with Topological Variance}

The quality of primary algorithm with respect to alternative algorithm is obtained by computation of Topological Variance. For e.g. the comparative performance of $AGDL$ for all the alternative algorithms is shown in Table~\ref{tvv}. Topological Variance of $AGDL$ concerning SCAN, LPA, GM on riskmap is 0.948, 0.956, and 0.956, respectively. Further,  we have obtained all such results to analyze the comparative performance of each primary algorithm with the alternative algorithms on the representative datasets. The results are represented using a heatmap, shown in Figure~\ref{ELA}.  Here, the  $X-$axis represents primary algorithms, $Y-$axis represents alternative algorithms and the right-hand side of $Y-$axis represents the $TV$ value, where dark green color indicates greater $TV$ score.  Now, if we consider the first heatmap where the left most, bottom most cell represented in dark green color gives the Topological Variance of $AGDL$ and $SCAN$. It indicates that $AGDL$ performs better as compared to $SCAN$. Similarly, as can be seen from the other cells in this column, $AGDL$ performs better than all the alternative algorithms as all the cells are dark green color.  Additionally, the performance of the other algorithms with respect to $AGDL$ is indicated in the top most row, which is light green color indicating poor performance. Therefore, these results suggests that $AGDL$ performs better than the alternative algorithms. Next, if we consider the performance of $Gdmp2$ for all the alternative algorithms and performance of alternative algorithms with respect to $Gdmp2$, then as can be seen from the respective cells, $Gdmp2$ gives worst performance.  Similarly, the results are plotted for all the algorithms on the remaining datasets.

\subsection{Result analysis with ranking}

Ranking is done to obtain a clear idea about the comparative performance of primary algorithms. The algorithm that gives the best quality community gets the highest rank (i.e. rank 1). Here, we consider the primary algorithm with the maximum Average Topological Variance as the best quality algorithm with the highest $DTV$ rank. The Average Topological Variance of all primary algorithms on all datasets is summarized in Table~\ref{elr}. It is interesting to notice that, on  riskmap dataset, the maximum Average Topological Variance is obtained by $AGDL$; hence, it gets the highest $DTV$ rank. Similarly, $AGDL$ obtains highest $DTV$ rank on Football, Jazz, Dolphin, Polbooks, Strike and Sawmill. Whereas, Kcut gets the highest $DTV$ rank on Polblogs. Results based on $DTV$ ranking is shown in Figure~\ref{quah}. In addition to tracking the performance of primary algorithms concerning all alternative algorithms, the performance on each of the datasets holds a significant role. Algorithms that give the maximum Average Topological Variance considering all datasets gets the highest $OTV$ rank. The $OTV$ rank of the baseline algorithms is summarized in Table~\ref{elr}. As the summation of Average Topological Variance obtained on all the datasets is maximum for $AGDL$, hence, $AGDL$ gets the highest $OTV$ rank. Following similar procedure, $OTV$ rank for the remaining algorithms are obtained.

\subsection{Result analysis with internal and external connections}
The primary algorithms are also analyzed based on the evaluation metrics such as; Isolability, Modularity and Conductance for all datasets. Here, we have summarized the results of the baseline algorithms on Dolphin and Polbooks shown in Table~\ref{rem}. The  results based on these three metrics suggests that $AGDL$ gives the best quality community and $Gdmp2$ gives worst quality community.

\section{Conclusion}
\label{six}
A direct quality comparison measure namely, \textsl{Topological Variance} is designed to evaluate community detection algorithms. This measure is used to compare the communities based on topological information. Two ranking schemes have been proposed based on \textsl{Topological Variance}. We have used eight real-world datasets and six community detection algorithms to perform comparative analysis of our proposed quality measure. The results given by our proposed measure and evaluation metrics affirm that $AGDL$ gives best quality communities and $Gdmp2$ gives worst quality communities indicating correctness of our proposed measure. In the future, we shall extend our measure to analyze the quality of communities at group level.




\end{document}